\documentclass[preprint,prb,showpacs,preprintnumbers,amsmath,amssymb]{revtex4}

\usepackage{graphicx}
\usepackage{dcolumn}
\usepackage{bm}

\begin{document}
\title
{
Significant suppression of weak ferromagnetism in (La${}_{1.8}$Eu${}_{0.2}$)CuO${}_4$
}
\author{I. Tsukada,
\email{ichiro@criepi.denken.or.jp} 
X. F. Sun, Seiki Komiya, A. N. Lavrov, and Yoichi Ando}
\affiliation{
Central Research Institute of Electric Power Industry, 
2-11-1 Iwado-kita, Komae, Tokyo 201-8511, JAPAN 
}
\date{\today}

\begin{abstract}
The magnetic structure of (La${}_{1.8}$Eu${}_{0.2}$)CuO${}_4$ has been 
studied by magnetization measurements of single crystals, 
which show antiferromagnetic long-range order below $T_N$ = 265~K 
and a structural phase transition at $T_s$ = 130~K. 
At $T_s < T < T_N$, the Cu spin susceptibility exhibits almost the same 
behavior as that of La${}_2$CuO${}_4$ in the low-temperature 
orthorhombic phase, which indicates the existence of finite spin canting 
out of the CuO${}_2$ plane. 
At $T < T_s$, the magnitude of the weak-ferromagnetic moment induced 
by the spin canting is suppressed approximately by 70{\%}. 
This significant suppression of the weak-ferromagnetic moment is 
carefully compared with the theoretical analysis of weak ferromagnetism 
by Stein {\it et al.} (Phys. Rev. B {\bf 53}, 775 (1996)), 
in which the magnitude of weak-ferromagnetic moments strongly depend on 
the crystallographic symmetry. 
Based on such comparison, 
below $T_s$ (La${}_{1.8}$Eu${}_{0.2}$)CuO${}_4$ is in the 
low-temperature less-orthorhombic phase with a space group of $Pccn$. 
We also discuss the possible magnetic structure of the pure low-temperature 
tetragonal phase with space group $P4_2/{ncm}$, 
which is relevant for rare-earth and alkaline-earth ions co-doped 
La${}_2$CuO${}_4$. 
\end{abstract}
\pacs{74.25.Ha,74.72.Dn,75.25.+z,75.30.Gw}

\maketitle


\section{Introduction}

\begin{figure}
\includegraphics*[width=75mm]{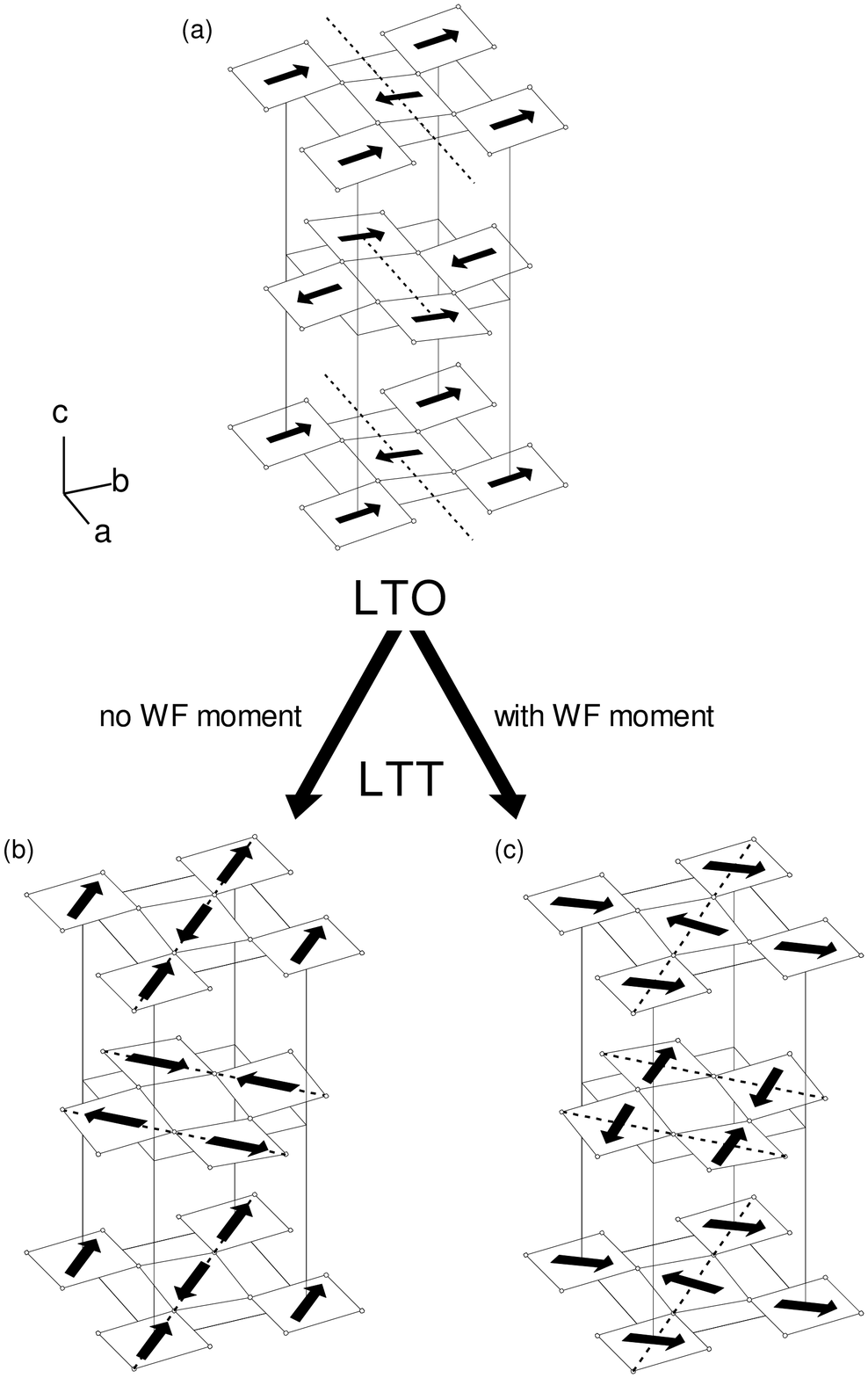}
\caption{(a) Magnetic structure of the LTO phase, 
where arrows show the spins on Cu${}^{2+}$ ions, 
and dashed lines show rotation axes of the CuO${}_6$ octahedra. 
(b,c) Two magnetic structures proposed for the LTT phase, 
without spin canting (b) and with spin canting (c). 
}
\label{Fig.1}
\end{figure}

The magnetic structure of La${}_2$CuO${}_4$ has been extensively studied 
because it is considered to be a good example of two-dimensional 
$S$=1/2 Heisenberg antiferromagnet,
whose nature is one of the central topics of quantum magnetism.
\cite{Chakravarty1} 
La${}_2$CuO${}_4$, however, shows a rather classical long-range N\'eel order 
at low temperatures because of its relatively low crystallographic symmetry 
caused by an orthorhombic distortion in the low-temperature orthorhombic 
(LTO) phase and a finite inter-plane coupling. 
In the LTO phase, spins are aligned approximately along the $b$ axis 
with slight canting out of the CuO${}_2$ plane,
\cite{Thio1,Vaknin1,Kastner1,Yamada1} 
which is responsible for the weak ferromagnetism of La${}_2$CuO${}_4$. 
The schematic drawing of the magnetic structure of La${}_2$CuO${}_4$ in the 
LTO phase is shown in Fig.~\ref{Fig.1} (a). 
The spin canting was simply explained
\cite{Coffey1,Coffey2} 
by antisymmetric exchange interactions of Dzyaloshinskii-Moriya (DM) type.
\cite{Dzyaloshinskii1,Moriya1} 
However, it has been pointed out 
by Shekthman {\it et al.},
\cite{SEA1,SEA2} 
Bonesteel,
\cite{Bonesteel1} 
and Yildirim {\it et al.}
\cite{Yildirim1,Yildirim2} 
that the true origin may not be so simple, because of the importance of 
an additional anisotropic symmetric spin-spin interaction. 
This interaction was already derived by Moriya
\cite{Moriya1} 
but unfortunately ignored for a long time. 
Kaplan was the first who noticed the importance of this interaction.
\cite{Kaplan1} 
He reinvestigated the role of spin-orbit interaction in super-exchange 
mechanism for one-dimensional $S$ = 1/2 antiferromagnetic (AF) spin chains 
and spin rings, 
and found that the anisotropic symmetric interaction gives an easy-axis 
anisotropy to the spin system that completely 
compensates the easy-plane anisotropy given by the DM interaction. 
Although the presence of this term (which we call KSEA interaction 
following Zheludev {\it et al.}\cite{Zheludev1}) 
has been recently established in several $S$ = 1/2 antiferromagnets, 
Ba${}_2$CuGe${}_2$O${}_7$,\cite{Zheludev1,Zheludev2} 
K${}_2$V${}_3$O${}_8$,\cite{Lumsden1} 
and Yb${}_4$As${}_3$,\cite{Shiba1} 
its role in the most interesting material La${}_2$CuO${}_4$ has not been 
experimentally revealed yet.

One possible test is to determine the magnetic structure of the so-called 
low-temperature tetragonal (LTT) phase (space group $P4_2/ncm$), 
which was first reported in (La,Ba)${}_2$CuO${}_4$.
\cite{Axe1}
Bonesteel
\cite{Bonesteel1} 
discussed that the weak-ferromagnetic (WF) moment per CuO${}_2$ 
plane should disappear in the LTT phase as is schematically drawn 
in Fig.~\ref{Fig.1} (b) when one takes the KSEA interaction into account. 
In this arrangement, spins are parallel to the rotation axes of 
CuO${}_6$ octahedra, and therefore, they are also parallel/antiparallel 
to the DM vectors. 
This theoretical prediction became a trigger of a long debate,
\cite{Koshibae1,Viertio1,Stein1} 
because the magnetic structure shown in Fig.~\ref{Fig.1} (b) was different 
from that proposed based on neutron diffraction experiments 
for (La,Nd)${}_2$CuO${}_4$ crystals.
\cite{Shamoto1,Crawford1,Keimer1} 
The latter magnetic structure is shown in Fig.~\ref{Fig.1} (c), 
where spins are perpendicular to the rotation axes of 
the CuO${}_6$ octahedra, and thus the spin canting remains intact. 
Since (La,Nd)${}_2$CuO${}_4$ was previously believed to be in the LTT phase, 
it was concluded that the LTT phase favored the spin arrangement 
shown in Fig.~\ref{Fig.1} (c) rather than that in Fig.~\ref{Fig.1} (b). 
By referring to the experimentally proposed model, 
Koshibae {\it et al.}
\cite{Koshibae1} 
tried to explain the presence of WF moment in the LTT phase.
They expanded the previously reported theories
\cite{SEA1,SEA2,Bonesteel1} 
by assuming the magnitude of DM and KSEA interactions independently, 
and predicted the presence of WF moment in the LTT phase 
for a particular situation where the effect of DM interaction is dominant. 
Stein {\it et al.}
\cite{Stein1} 
also calculated the WF moment with similar assumptions as Koshibae {\it et al.}
\cite{Koshibae1} 
do, 
but concluded that the physically reasonable ratio of DM and KSEA 
interactions do not favor the presence of WF moment in the pure LTT phase. 
Therefore, the above mentioned controversy among theories and 
neutron experiments has not been solved yet.

In this paper we study the anisotropic magnetization of 
(La${}_{1.8}$Eu${}_{0.2}$)CuO${}_4$ single crystals and discuss 
the magnetic structure to shed light on this problem of long debate. 
Since Eu${}^{3+}$ ions do not give large Curie-type susceptibility 
in contrast to Nd${}^{3+}$, 
(La${}_{1.8}$Eu${}_{0.2}$)CuO${}_4$ is a more suitable 
system for magnetization measurements. 
Although magnetic susceptibility data of polycrystalline samples 
have been already reported by Kataev {\it et al.},
\cite{Kataev1} 
one needs the $c$-axis susceptibility data measured on a single crystal 
to evaluate the magnitude of spin canting out of the CuO${}_2$ plane. 
We observe a significant suppression of the spin canting 
below the structural phase-transition temperature, 
which has not yet been elucidated to be a transition from the LTO to the LTT 
phase, or to the low-temperature less orthorhombic 
(LTO2, space group $Pccn$) phase. 
We analyze the change of WF moment across the structural phase transition, 
and compare the magnitude of the spin canting with 
that predicted by Stein {\it et al.}
\cite{Stein1}; 
the obtained results point to the LTO2 crystallographic symmetry of 
(La${}_{1.8}$Eu${}_{0.2}$)CuO${}_4$ below $T_s$. 
Finally we will discuss the likely magnetic structure for pure LTT symmetry, 
and make some comments on the previous controversy in understanding 
the weak ferromagnetism in La${}_{1.65}$Nd${}_{0.35}$CuO${}_4$.

\section{Experimental results}

(La${}_{1.8}$Eu${}_{0.2}$)CuO${}_4$ single crystals are grown 
by a traveling-solvent floating zone method,
\cite{Komiya1} 
and their quality is confirmed by x-ray diffractions. 
The crystalline rod is carefully cut into a rectangular shape 
with all the faces normal to the principal axes, which is confirmed 
using an x-ray Laue backscattering. 
Then the samples are annealed in He-gas flow to remove excess oxygen. 
Magnetic susceptibility measurements are carried out with commercial SQUID 
magnetometer (MPMS, Quantum Design) in magnetic field up to $H$ = 70~kOe. 
To avoid confusion, we follow the axis notation of the LTO phase, 
so that the $a$ and $b$ axes run along two orthogonal in-plane Cu-Cu diagonal 
directions and the $c$ axis is normal to the CuO${}_2$ plane.

\begin{figure}
\includegraphics*[width=75mm]{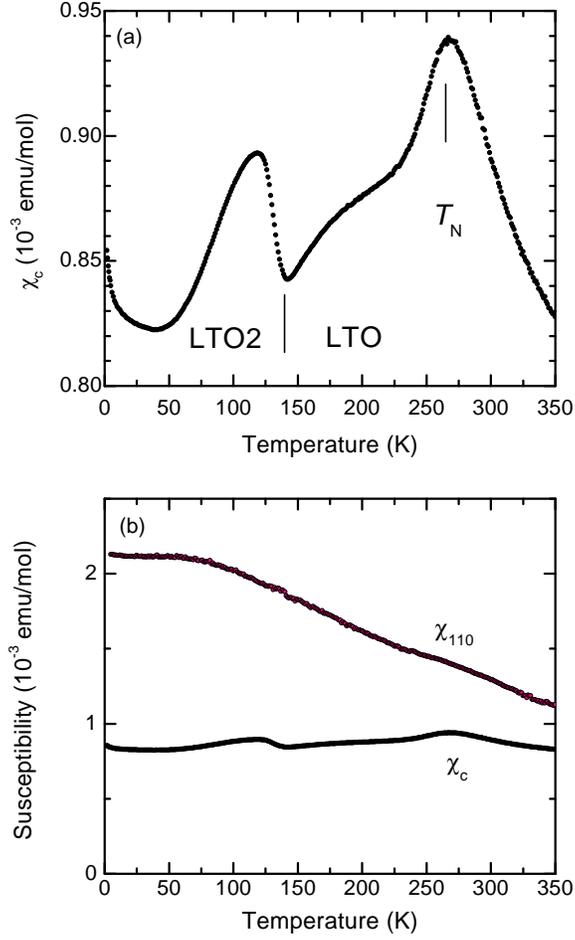}
\caption{(a) Temperature dependence of $\chi_c$ of 
(La${}_{1.8}$Eu${}_{0.2}$)CuO${}_4$ at $H$ = 5~kOe. 
The structural phase-transition temperature $T_s$ is 
taken from published data.\protect\cite{Suh1,Klauss1}
(b) A comparison of $\chi_c$ and $\chi_{110}$ at $H$ = 5~kOe. 
}
\label{Fig.2}
\end{figure}

Figure~\ref{Fig.2} (a) shows the magnetic susceptibility of 
(La${}_{1.8}$Eu${}_{0.2}$)CuO${}_4$ measured along the $c$ axis 
($\chi_c$) at $H$ = 5~kOe. 
The peak structure is identified as a N{\'e}el transition at $T_N$ = 265~K 
similar to that of Eu-free La${}_2$CuO${}_4$.
\cite{Thio1,Lavrov1} 
In addition to this peak, we observe a rapid increase in $\chi_c$ below 
$T$ = 130~K as was reported for polycrystalline samples.
\cite{Kataev1} 
Before discussing $\chi_c$ in detail, we should consider the contribution 
from Eu${}^{3+}$ ions. 
The magnitude of $\chi_c$ of (La${}_{1.8}$Eu${}_{0.2}$)CuO${}_4$ 
is several times larger than that of Eu-free La${}_2$CuO${}_4$,
\cite{Thio1,Lavrov1} 
which is likely to be due to the Van-Vleck contribution of Eu${}^{3+}$ ions. 
This Van-Vleck contribution is anisotropic as can be seen from the data of 
$\chi_c$ and $\chi_{110}$ (susceptibility measured along the [110] direction) 
shown in Fig.~\ref{Fig.2} (b). 
$\chi_{110}$ exhibits a large-scale increase upon cooling from room 
temperature to 80~K similar to that observed in Eu${}_2$CuO${}_4$.
\cite{Tovar1} 
A slight and broad hump around 265~K in $\chi_{110}$ comes from the N{\'e}el 
transition. On the other hand, such large-scale increasing behavior upon 
cooling is not found in $\chi_c$ of (La${}_{1.8}$Eu${}_{0.2}$)CuO${}_4$ 
in contrast to $\chi_c$ of Eu${}_2$CuO${}_4$ 
where the Van-Vleck contribution of Eu${}^{3+}$ ions gives a similar increase 
of $\chi_c$ also down to approximately 150~K.
\cite{Tovar1} 
We speculate that the different oxygen coordinations around Eu${}^{3+}$ ion, 
rock-salt type in (La${}_{1.8}$Eu${}_{0.2}$)CuO${}_4$  and fluorite type 
in Eu${}_2$CuO${}_4$, are the origin of the different Van-Vleck contributions. 
In the following discussion, we assume that the temperature dependence 
of $\chi_c$ is mostly coming from the magnetic response of CuO${}_2$ planes.

Our sample shows somewhat lower $T_N$ in comparison with 
polycrystalline (La${}_{1.8}$Eu${}_{0.2}$)CuO${}_4$ (Ref. [27]) 
or single-crystalline La${}_2$CuO${}_4$ (Ref. [29]), 
which suggests that a small amount of excess oxygen may remain in our sample. 
However, this does not change the magnetic structure of AF long-range order, 
and we can safely neglect its effect in the present context. 
Below $T_N$, $\chi_c$ decreases down to $T_s$ = 130~K, 
which coincides with the reported LTO $\rightarrow$ LTO2/LTT structural 
phase-transition temperature
\cite{Suh1,Klauss1} 
and then steeply increases until $T$ = 110~K; 
such increase is not seen in $\chi_{110}$. 
The size of this increase is approximately 5 $\times$ 10${}^{-5}$~emu/mol. 
This is larger than that observed in polycrystalline samples,
\cite{Kataev1} 
which is apparently because of averaging of $\chi_{110}$ and $\chi_c$. 
This increase at $T_s$ is most likely related to the change of inter-plane 
magnetic couplings as will be discussed later.

\begin{figure}
\includegraphics*[width=75mm]{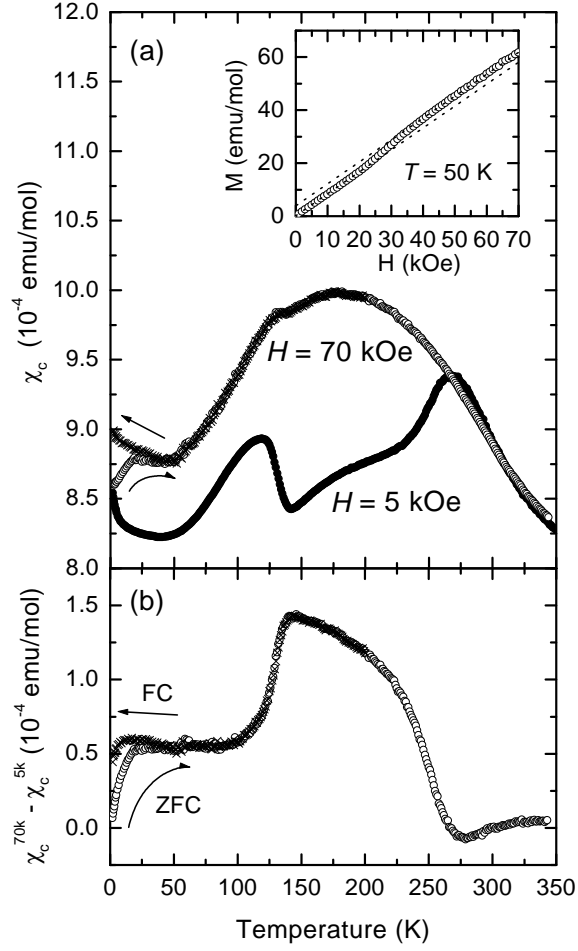}
\caption{(a) Temperature dependence of $\chi_c$ at magnetic fields 
below $H_c$ (5~kOe) and above $H_c$ (70~kOe). 
$\chi_c$ at $H$ = 70~kOe ($\chi_c^{70{\rm k}}$) shows a hysteresis between 
ZFC and FC data below 20~K. 
Inset to (a): Field dependence of the magnetization along the $c$ axis 
at $T$ = 50~K. 
Two dotted lines are linear fits to the low- and high-field 
magnetizations. 
(b) Temperature dependence of $\chi_c^{70{\rm k}}$ - $\chi_c^{5{\rm k}}$. 
}
\label{Fig.3}
\end{figure}

In order to clarify whether the magnitude of WF moment is changed across the 
structural phase transition, we measure the high-field susceptibility, 
which is shown in Fig.~\ref{Fig.3} (a). 
The temperature dependence of $\chi_c$ at $H$ = 70~kOe ($\chi_c^{70{\rm k}}$) 
is notably different from that at $H$ = 5~kOe ($\chi_c^{5{\rm k}}$) 
below $T_N$; 
it increases continuously instead of forming a peak structure at $T_N$, 
shows a broad maximum around 170~K, and then decreases toward 50~K. 
Below 20~K there is a hysteresis between zero-field cool (ZFC) and field cool 
(FC) data as conventional ferromagnets with a finite coercive force show, 
which suggests the presence of magnetic domains. 
It should be noted that such a hysteresis is absent in La${}_2$CuO${}_4$, 
and only ligtly hole-doped (La${}_{2-x}$Sr${}_x$)CuO${}_4$ exhibits 
hysteretic behavior between FC and ZFC process.
\cite{Lavrov1} 
It is likely that the substitution of Eu${}^{3+}$ for La${}^{3+}$ induces 
a structural disorder that causes a domained structure in our sample.

The field dependence of the magnetization at $T$ = 50~K [inset of 
Fig.~\ref{Fig.3} (a)] clearly shows the WF transition around $H_c$ = 30~kOe. 
The transition width is large in comparison with that of La${}_2$CuO${}_4$.
\cite{Thio2}
These data give us two important clues to reveal the magnetic 
structure of (La${}_{1.8}$Eu${}_{0.2}$)CuO${}_4$. 
One is a finite but significantly suppressed WF moment and the other is 
a decrease of the critical field from that of La${}_2$CuO${}_4$.
\cite{Lavrov2} 
To see the suppression of WF moment more clearly, we plot 
$\Delta\chi_c$ = $\chi_c^{70{\rm k}}$ - $\chi_c^{5{\rm k}}$, 
which is shown in Fig.~\ref{Fig.3} (b). 
Below $T_N$, $\Delta\chi_c$ follows the temperature dependence 
typical for the evolution of spontaneous magnetization. 
Just above $T_s$, $\Delta\chi_c$ reaches almost 
1.5 $\times$ 10${}^{-4}$~emu/mol, and its extrapolation to the lower 
temperatures gives $\Delta\chi_c$ $\approx$ 1.7 $\times$ 10${}^{-4}$~emu/mol 
at $T$ = 0~K. 
This value corresponds to the spontaneous magnetization of 12~emu/mol, 
which would be the WF moment if the LTO structure were maintained 
down to $T$ = 0~K. 
In the real sample, however, the structural phase transition occurs, 
and $\Delta\chi_c$ is reduced approximately to 5 $\times$ 10${}^{-5}$ emu/mol 
($M_{WF}$). 
This value gives the magnetization of 3.5~emu/mol, consistent with the vertical 
shift of two dotted lines in the inset of Fig.~\ref{Fig.3} (a). 
Therefore, the magnitude of WF moment is suppressed almost by 70{\%}. 
Moreover, the decrease of the critical field indicates a reduction 
in the inter-plane coupling. 
The critical field of the WF transition is mainly determined 
by the competition between AF inter-plane coupling and Zeeman effect 
$H_c$ $\times$ $M_{WF}$ which is lowered approximately by 5 times. 
Thus, the AF inter-plane coupling is reduced as well.

\section{Discussion}

\subsection{Theoretical predictions of the magnetic structure}

Let us briefly survey several theoretical works on the magnetic structure of 
rare-earth doped La${}_2$CuO${}_4$. 
In order to analyze the experimental data, we need information 
of the WF moment not only in the LTO and LTT phases but also in the LTO2 phase. 
To our knowledge, Coffey {\it et al.}
\cite{Coffey2} 
was the first who discussed possible 
magnetic structures of La${}_2$CuO${}_4$ in the LTT phase. 
They proposed two kinds of ground state: 
one with a canted moment per CuO${}_2$ plane and one without. 
After Shekhtman {\it et al.}
\cite{SEA1} 
pointed out the importance of the KSEA term upon discussing the magnetic 
structure of LTO phase, 
Bonesteel
\cite{Bonesteel1} 
tried to include this term in calculations of the magnetic ground state 
of the LTT phase, 
and concluded that the canted moment per CuO${}_2$ plane should be zero 
(Fig.1(b)) if the KSEA term is properly incorporated into the spin Hamiltonian. 
In the Bonesteel's paper, a structural difference between 
the LTO2 and LTT phases has been carefully considered; 
a spin canting angle out of the CuO${}_2$ plane has been derived to be 
proportional to the angle $\kappa$ that is the angle between the Cu-O-Cu bond 
direction and the rotation axis of CuO${}_6$ octahedron 
(${\kappa}={\pi}/4$, $0<{\kappa}<{\pi}/4$, and ${\kappa}=0$, 
for the LTO, LTO2, and LTT phases, respectively). 
Vierti{\"o} and Bonesteel
\cite{Viertio1} 
then discussed the relation between the interplanar couling and 
the weak ferromagnetic transition in (La${}_{2-x}$Nd${}_x$)CuO${}_4$ 
using the same assumptions as in Ref. 12.

The theoretical calculation done by Stein {\it et al.}
\cite{Stein1} 
is basically the same as that done by Bonesteel and co-workers
\cite{Bonesteel1,Viertio1}, 
but they have evaluated the WF moment more rigorously.
\cite{Stein1} 
They calculated the $\kappa$ dependence 
(it is denoted as $\alpha$ in their original paper) of the WF moment 
under several different conditions, 
which is reproduced in Fig.~\ref{Fig.4}. 
The vertical axis $M_F/(SD/2J)$ is the WF moment normalized 
by the magnitude of spin, Heisenberg interaction and DM interaction. 
The only tunable parameter, $\Omega$, qualitatively gives the relative 
magnitude of the DM and KSEA interactions; 
$\Omega$ = 1 corresponds to the situation without KSEA interaction 
(Heisenberg + DM), and $M_F/(SD/2J)$ = 1 regardless of $\kappa$, 
while $\Omega$ = 0 corresponds to the situation where the KSEA interaction 
completely compensates the easy-plane anisotropy given by the DM interaction 
as was derived by Bonesteel
\cite{Bonesteel1} 
(Heisenberg + DM + KSEA). 
Intermediate values of $\Omega$ correspond to the situation that the 
easy-plane anisotropy caused by the DM interactions still survives 
even under the presence of KSEA interactions. 
Since the case discussed by Koshibae {\it et al.} is included in this 
parameter region, 
Stein's calculation may reproduce all the possible situations by tuning 
$\Omega$, and thus, it is most convenient for analyzing our experimental 
data.

\subsection{Magnetic structure of (La${}_{1.8}$Eu${}_{0.2}$)CuO${}_4$}

\begin{figure}
\includegraphics*[width=75mm]{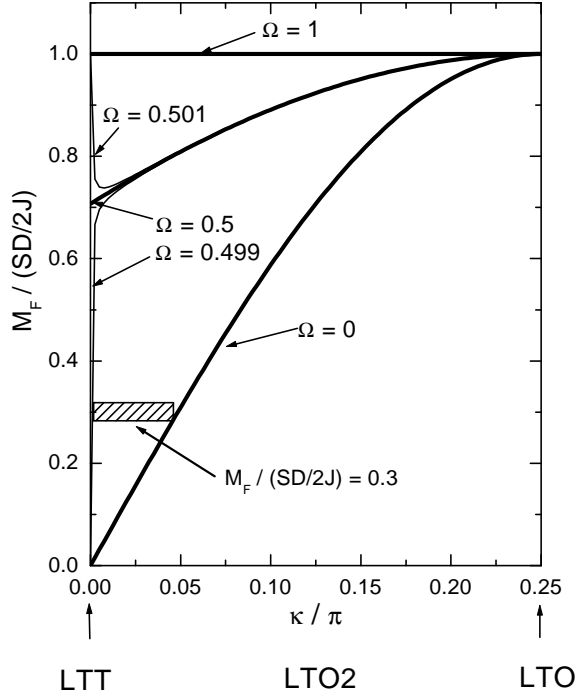}
\caption{
Theoretical calculation of the WF moment by Stein {\it et al.}
\protect\cite{Stein1} 
Horizontal axis is the angle between the Cu-O-Cu bond direction and 
the rotation axis of CuO${}_6$ octahedra: 
$\kappa$ = 0, 0 ${<} {\kappa} {<}$ $\pi$/4, and $\kappa$ = $\pi$/4 
correspond to the LTT, LTO2, and LTO phases, respectively. 
Vertical axis is the WF moment normalized by the magnitude of spin, 
Heisenberg interaction, and DM interaction. 
The only tunable parameter, $\Omega$, qualitatively gives the relative 
magnitude of the DM and KSEA interactions. 
For any $\Omega$ values, the WF moment in the LTO2 phase is smaller than 
that in the LTO phase. 
$\Omega$ = 0.5 line gives the boundary: 
The WF moment completely disappears in the LTT phase when $\Omega <$ 0.5, 
while the system recovers the WF moment in the LTT phase 
when $\Omega >$ 0.5. 
The line for $\Omega$ = 0 gives the situtaion that formerly discussed by 
Bonesteel.
\protect\cite{Bonesteel1}
The shaded band shows the position of the WF moment observed in this work. 
}
\label{Fig.4}
\end{figure}

Based on the Stein's calculations, we can discuss the source of 
the 70{\%} suppression of the WF moment across $T_s$. 
Figure~\ref{Fig.4} teaches us that the WF moment would be the same in the LTO, 
LTO2, and LTT phases when only the DM interaction is effective. 
This is however inconsistent with our observation. 
We do observe the suppression of the WF moment, 
and this fact itself indicates that the KSEA interaction actually works in 
(La${}_{1.8}$Eu${}_{0.2}$)CuO${}_4$. 
It should be noted that such quantitative comparison of the WF moment 
across $T_s$ can be done only by magnetization measurements, 
while a neutron diffraction measurement is not sensitive enough to 
judge whether the spin canting is present or not. 
Thus, it is not surprising that previous neutron experiments failed to 
detect this change.

Figure~\ref{Fig.4} also gives the range of $\Omega$ and $\kappa$ 
for our sample. 
One can easily notice that the critical line, $\Omega = 0.5$, is terminated 
at $M_F/(SD/2J)$ = 1/{$\sqrt{2}$} $\approx$ 0.70 at $\kappa$ = 0. 
This means that if one observes the WF moment less than 70{\%} of that 
in the LTO phase, $\Omega$ must be smaller than 0.5. 
The WF moment of (La${}_{1.8}$Eu${}_{0.2}$)CuO${}_4$ below $T_s$ is only 
$\approx$ 30{\%} of that just above $T_s$ 
as we indicate by the position of $M_F/(SD/2J)$ $\approx$ 0.3 
in Fig.~\ref{Fig.4}. 
Therefore, to produce such small WF moment, 
$\kappa$ should be limited to the region 0 $<$ $\kappa$ $<$ $\pi$/20. 
This suggests that our sample is in the LTO2 phase even though 
the rotation angle is close to that for the LTT phase.

It should be pointed out that our identification of the structural symmetry 
is different from the reported symmetry for (La,Eu)${}_2$CuO${}_4$,
\cite{Suh1,Klauss1} 
which was suggested to be the LTT phase at the lowest temperature. 
However, to our knowledge, there has been no conclusive structural analysis 
for the low-temperature phase of (La,Eu)${}_2$CuO${}_4$. 
Since the symmetry below $T_s$ was not important in many studies on 
(La,Eu)${}_2$CuO${}_4$, peoples have not paid attention to the difference 
between the LTO2 and LTT phases. 
What we found in the present work is that the spin canting 
out of the CuO${}_2$ plane is quite sensitive to the structural difference 
between these phases 
that is the reason why we can distingish the LTO2 phase from the LTT phase. 
We also emphasize that $\Omega$ is definitely smaller than 0.5 for 
(La,Eu)${}_2$CuO${}_4$. 
This is consistent with the conclusion by Stein {\it et al.},
\cite{Stein1} 
where they estimated $\Omega$ to be 0.10 from several microscopic parameters.

\begin{figure}
\includegraphics*[width=50mm]{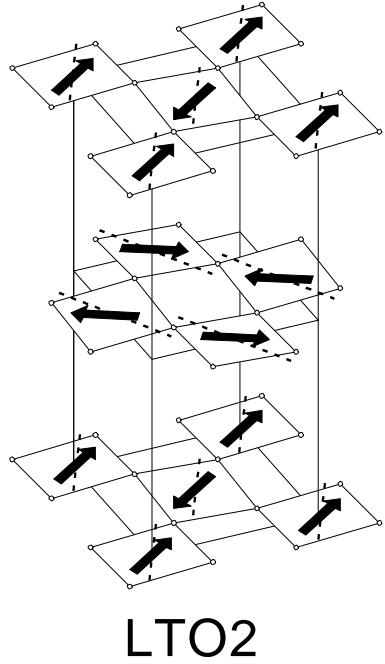}
\caption{Magnetic structure for the LTO2 phase of 
(La${}_{1.8}$Eu${}_{0.2}$)CuO${}_4$, 
which is considered as an intermediate state 
between the LTO (Fig.~1(a)) and the LTT (Fig.1~(b)) phases. 
In this state, a finite spin canting remains, but its angle is largely 
suppressed from that in the LTO phase. 
}
\label{Fig.5}
\end{figure}

As a result, we arrive at a conclusion that the most probable magnetic 
structure for LTO2 (La${}_{1.8}$Eu${}_{0.2}$)CuO${}_4$ should be 
as shown in Fig.~\ref{Fig.5}, 
where spins are parallel to neither [100] nor [110] directions. 
This model implies a reduction of the effective inter-plane coupling, 
which explains the increase of $\chi_c$ below $T_s$. 
The inter-plane coupling is determined by a subtle balance of the exchange 
interactions mainly between spins at (0, 0, 0) and ($a/2$, 0, $c/2$) 
and that between spins at (0, 0, 0) and (0, $b/2$, $c/2$). 
Since the above two interactions are not equivalent in the LTO phase
\cite{Uchinokura1}, 
the effective inter-plane coupling remains finite. 
This inequivalence should decrease according to the symmetry change from 
the LTO to the LTO2 phases, and finally disappear in the LTT phase. 
The effective coupling between spins at the neighboring layers then 
decreases, and spins become more susceptible to the magnetic field 
applied along the $c$ axis. 
A more detaied discussion on the effect of inter-plane coupling 
has been already given by Viertio and Bonesteel.
\cite{Viertio1} 
It is most likely that the reduction of the effective inter-plane coupling 
is the main reason why $\chi_c$ increases below $T_s$. 
It should be noted that the Van-Vleck susceptibility of Eu${}^{3+}$ ions 
can also contribute to the rapid increase of $\chi_c^{5{\rm k}}$ below $T_s$, 
because the oxygen coordination around Eu${}^{3+}$ ions is modified 
across this temperature. 
However, this effect is sufficiently small as is suggested by the $c$-axis 
susceptibility data of (La${}_{1.65}$Eu${}_{0.2}$Sr${}_{0.15}$)CuO${}_4$,
\cite{Simovic1}
and, therefore, one may safely attribute the large increase of 
$\chi_c^{5{\rm k}}$ below $T_s$ to the response from CuO${}_2$ planes.

\subsection{Implication for the magnetic structure in the LTT phase}

Finally we briefly discuss the likely magnetic structure of the LTT phase. 
Our results indicate that $\Omega$ $<$ 0.5. 
This parameter implies that the WF moment should disappear in the LTT phase 
($\kappa$ = 0). 
Thus the magnetic structure of the LTT phase is expected to be consistent 
with that shown in Fig.~\ref{Fig.1} (b), 
but is inconsistent with that shown in Fig.~\ref{Fig.1} (c).  
A natural next step is to find a system that exhibits a true LTT symmetry 
without doped carriers, 
in which we may expect a perfect disappearance of the effective 
inter-plane couplings. 
Unfortunately, suitable compound possessing a true LTT symmetry and 
AF long-range order has not been found yet. 
However, we believe that we can find a better system other than 
(La,Eu)${}_2$CuO${}_4$ in near future, which will be useful for the 
study of pure two-dimensional $S$=1/2 AF spin systems.

\section{Summary}

To summarize, we have investigated the magnetic susceptibility 
of (La${}_{1.8}$Eu${}_{0.2}$)CuO${}_4$.  
The antiferromagnetic long-range order typical for the LTO phase 
establishes below $T_N$ = 265~K, 
where a finite spin canting out of the CuO${}_2$ plane appears. 
In contrast to Eu-free La${}_2$CuO${}_4$, (La${}_{1.8}$Eu${}_{0.2}$)CuO${}_4$ 
shows an additional structural phase transition at $T_s$ = 130~K, 
below which the spin canting is suppressed by 70{\%}. 
All the results are consistently explained by the theory given by 
Stein {\it et al.,}
\cite{Stein1} 
in which the KSEA interactions are taken into account together with the DM 
interactions. 
According to this theory, the spin arrangement for the LTO2 phase is 
determined. 
Furthermore, we observe a reduction of the effective inter-plane coupling 
below the structural transition temperature, which is probably one of the 
keys to understand the various anomalies found in the LTT phase of 
(La,Nd,Sr)${}_2$CuO${}_4$.

\section*{Acknowledgment}

The authors thank M. Matsuda and W. Koshibae for fruitful discussions.

\end{document}